\documentclass[aps,prl,preprintnumbers,amsmath,amssymb,twocolumn,superscriptaddress,showpacs,floatfix]{revtex4-2}
\usepackage[dvips,final]{graphicx}
\usepackage{pstricks}
\usepackage{epsf}
\usepackage{amsmath}
\usepackage{amsfonts}
\usepackage{amssymb}
\usepackage{dcolumn}
\usepackage{bm}
\usepackage{mathrsfs}
\usepackage{slashed}
\usepackage{xcolor}

\usepackage{mathrsfs}

\usepackage[breaklinks, colorlinks=true, citecolor=blue, linkcolor=cyan]{hyperref}

\begin{document}


\title{Weyl anomaly induced transport in hydrodynamics}

\author{Shi-Zheng Yang}
\email{yangsz@ustc.edu.cn}
\affiliation{School of Information Engineering, Zhejiang Ocean University, Zhoushan, Zhejiang 316022, China}
\affiliation{Department of Modern Physics and Anhui Center for fundamental Sciences
(Theoretical Physics), University of Science and Technology of China,
Anhui 230026}

\author{Jian-Hua Gao}
\email{gaojh@sdu.edu.cn}
\affiliation{Shandong Provincial Key Laboratory of Nuclear Science, Nuclear Energy Technology and Comprehensive Utilization,
Weihai Frontier Innovation Institute of Nuclear Technology, School of Nuclear Science, Energy and Power Engineering, Shandong University, Shandong 250061, China}
\affiliation{Weihai Research Institute of Industrial Technology of Shandong University, Weihai
264209, China}

\author{Zuo-Tang Liang}
\email{liang@sdu.edu.cn}
\affiliation{Institute of Frontier and Interdisciplinary Science,
Key Laboratory of Particle Physics and Particle Irradiation (MOE),
Shandong University, Qingdao, Shandong 266237, China}

\author{Georgy~Yu.~Prokhorov}
\email{prokhorov@theor.jinr.ru}
\affiliation{Joint Institute for Nuclear Research, Joliot-Curie 6, Dubna, 141980, Russia}
\affiliation{NRC Kurchatov Institute, Moscow, Russia}

\author{Shi Pu}
\email{shipu@ustc.edu.cn}
\affiliation{Department of Modern Physics and Anhui Center for fundamental Sciences
(Theoretical Physics), University of Science and Technology of China,
Anhui 230026}
\affiliation{Southern Center for Nuclear-Science Theory (SCNT), Institute of Modern Physics,
Chinese Academy of Sciences, Huizhou 516000, Guangdong, China}

\author{Oleg V. Teryaev}
\email{teryaev@jinr.ru}
\affiliation{Joint Institute for Nuclear Research, Joliot-Curie 6, Dubna, 141980, Russia}
\affiliation{NRC Kurchatov Institute, Moscow, Russia}

\author{Valentin I. Zakharov}
\email{vzakharov@itep.ru}
\affiliation{NRC Kurchatov Institute, Moscow, Russia}
\affiliation{Joint Institute for Nuclear Research, Joliot-Curie 6, Dubna, 141980, Russia}


\begin{abstract}
We show that the Weyl (trace) anomaly gives rise to a new non-dissipative vector current in accelerated relativistic fluids. The anomaly uniquely fixes the second-order transport coefficient governing the coupling between the electromagnetic field and the fluid acceleration. We derive this result by extending hydrodynamic anomaly matching to include the trace anomaly, and independently reproduce it in boundary quantum field theory by treating the Rindler horizon of an accelerated observer as an effective boundary. From the boundary perspective, the electric- and magnetic-field sectors correspond to screening and vacuum magnetization effects near the boundary. In the local rest frame, the electric-field contribution induces an additional charge density, while the magnetic-field contribution generates a transverse current with a Nernst-like, more generally thermomagnetic Hall-like, tensor structure. Our results reveal a new class of anomaly-induced transport governed by the trace anomaly.
\end{abstract}

\maketitle


\emph{Introduction}---Quantum anomalies provide a direct link between microscopic quantum field theory and macroscopic transport in relativistic many-body systems. In thermodynamic equilibrium, anomaly coefficients can generate non-dissipative contributions to currents, thereby encoding ultraviolet quantum information into infrared response. Canonical examples are the chiral magnetic effect (CME) \cite{Vilenkin:1980fu,Fukushima:2008xe,Kharzeev:2009pj,Kharzeev:2013jha}, which induces an electric current along an external magnetic field in the presence of chiral imbalance, and the chiral vortical effect (CVE) \cite{Vilenkin:1980zv,Son:2009tf,Landsteiner:2011cp}, which generates a current along the vorticity of a rotating medium. Related anomaly-induced transport phenomena have also been explored in broader settings, including the kinematical vortical effect \cite{Prokhorov:2022udo,Prokhorov:2026swu} and other extensions \cite{Gao:2012ix,Chen:2012ca, Zakharov:2012vv,Sadofyev:2010is,Stone:2018zel,Yang:2022ksq,Neiman:2010zi,Golkar:2012kb}, also see Refs.~\cite{Kharzeev:2024zzm,Kharzeev:2020jxw,Hidaka:2022dmn} for recent reviews. By contrast, whether the Weyl (trace) anomaly can directly induce universal macroscopic transport remains much less understood.

The Weyl anomaly reflects the quantum breaking of classical conformal invariance and manifests itself through a non-vanishing trace of the energy-momentum tensor. In the presence of a background electromagnetic field, it takes the form
\begin{eqnarray}
{T}^{\mu}_{\mu} &=& C F^{\mu\nu}F_{\mu\nu},,
\label{eq:anom}
\end{eqnarray}
with $C$ fixed by the field content of the theory. Unlike the chiral anomaly, this relation constrains the energy-momentum sector rather than directly sourcing a non-conserved current, making its transport consequences much less transparent. As a result, existing discussions of Weyl-anomaly-induced transport have been largely restricted to special settings \cite{Basar:2012bp,Chu:2018ksb,Chernodub:2021nff,Chernodub:2016lbo,Chernodub:2017jcp}, while its role in relativistic hydrodynamics, in particular whether it can fix universal transport, remains unclear.

A further clue comes from boundary quantum field theory (BQFT) \cite{Cardy:2004hm}, where the Weyl anomaly induces a current near a boundary in the presence of an external magnetic field, interpreted as a vacuum magnetization current \cite{Chu:2018ksb,Chu:2018ntx, Chu:2019rod}. This suggests that the anomaly-induced response is closely tied to boundary structure rather than to ordinary transport in matter. It is then natural to ask whether an analogous effect can arise in an accelerated frame, where the Rindler horizon provides an effective boundary. Establishing such a connection would provide a bridge between the boundary perspective of BQFT and a macroscopic hydrodynamic description of Weyl-anomaly-induced transport.

In this work, we fill this gap by showing that the Weyl anomaly induces a nondissipative vector current, which emerges in relativistic hydrodynamics at global equilibrium and is independently reproduced in boundary quantum field theory. More specifically, the anomaly uniquely fixes a second-order transport coefficient associated with the coupling between fluid acceleration and an external electromagnetic field. We establish this result in two independent ways. First, by extending the anomaly-matching logic of Son and Surowka \cite{Son:2009tf} and its modern equilibrium formulation \cite{Yang:2022ksq} to the trace anomaly in slowly varying external electromagnetic fields, we show that global-equilibrium constraints together with the anomalous trace uniquely determine the allowed constitutive structure. Second, we reproduce the same current in BQFT by treating the Rindler horizon of an accelerated observer as an effective boundary. The agreement between the two derivations reveals a nontrivial link between boundary-induced response and Weyl-anomaly-driven hydrodynamic transport. 
These results point to a new class of anomaly-induced transport governed by the trace anomaly, with the corresponding transport coefficient fixed by the Weyl anomaly. 
Throughout this work, we adopt the metric $g^{\mu\nu}=\{+,-,-,-\}$ and Levi-Civita tensor $\epsilon^{0123}=+1$.


\vspace{0.5cm}
\emph{Weyl-anomaly-induced currents from hydrodynamics}---
We first formulate the equilibrium setup and then show that the Weyl anomaly uniquely determines one of the second-order transport coefficients. Our starting point is the conservation laws for the energy-momentum tensor $T^{\mu\nu}$ and current $j^\mu$, 
\begin{eqnarray}
\partial_{\mu}T^{\mu\nu}=F^{\nu\mu}j_{\mu},\qquad \partial_{\mu}j^{\mu}=0.
\label{conservation}
\end{eqnarray}
The electromagnetic field-strength tensor can be decomposed as $F_{\mu\nu}=E_{\mu}u_{\nu}-E_{\nu}u_{\mu}+\varepsilon_{\mu\nu\alpha\beta}u^\alpha B^\beta$. 
For simplicity, we set the magnetic field $B^\mu$ to zero. The case with nonvanishing magnetic fields will be discussed later. 
The Weyl anomaly \eqref{eq:anom} then takes the simple form
\begin{equation}
g_{\mu\nu}T^{\mu\nu}=2C E^{2}.
\label{eq:trace_hydro}
\end{equation}

\vspace{0.2cm}
\noindent
\textbf{Hydrodynamic at global equilibrium:}
To identify possible novel nonvanishing transport effects induced by the trace anomaly, we focus on global equilibrium. At global equilibrium, the system satisfies the Killing conditions \cite{DeGroot:1980dk, Becattini:2014yxa}, 
\begin{eqnarray}
\partial_{\mu}\beta_{\nu}+\partial_{\nu}\beta_{\mu} & = & 0,\label{eq:Killing_eq_01}\\
\partial_{\mu}\overline{\mu}+F_{\mu\nu}\beta^{\nu} & = & 0,\label{eq:Killing_eq_02}
\end{eqnarray}
where the inverse-temperature vector $\beta_{\mu}\equiv u_{\mu}/T$ is a Killing vector and $\overline{\mu}\equiv\mu/T$.
Eq.~(\ref{eq:Killing_eq_01}) further implies that the thermal vorticity $\varpi_{\mu\nu}=(\partial_{\nu}\beta_{\mu}-\partial_{\mu}\beta_{\nu})/2$ is constant at global equilibrium. 
Similar to $F^{\mu\nu}$, we set the magnetic component of the thermal vorticity to zero. We then have $T\varpi_{\mu\nu}=a_{\mu}u_{\nu}-a_{\nu}u_{\mu}$,
where the electric component or acceleration part of the thermal vorticity is defined by $a^{\mu}=-T\varpi^{\mu\nu}u_{\nu}$. Using $\partial_{\rho}\varpi_{\mu\nu}=0$, we further obtain
\begin{equation}
a_{\mu}=u^{\nu}\partial_{\nu}u_{\mu}=T^{-1}\partial_{\mu}T,
\label{eq:a_def}
\end{equation}
which shows that the fluid acceleration $a_{\mu}$ becomes equal to the normalized temperature gradient at the global equilibrium. Acting with $\partial_{\nu}$ on Eq.~(\ref{eq:Killing_eq_02}) and using the commutativity of partial derivatives, one finds \cite{Yang:2024sfp}
\begin{equation}
u^{\lambda}(\partial_{\lambda}F_{\mu\nu})=a_{\nu}E_{\mu}-a_{\mu}E_{\nu}.
\label{eq:aE_01}
\end{equation}
This relation connects the electric field to the fluid acceleration. 

\vspace{0.2cm}
\noindent
\textbf{Mapping to the Weyl anomaly:}
We next perform a gradient expansion for the system and expand the current $j^{\mu}$ and the energy-momentum tensor $T^{\mu\nu}$ order by order. The leading- and first-order structures are well known; see, for example, the early pioneering work \cite{Israel:1979wp}, the recent review \cite{Rocha:2023ilf}, and the references therein.. In the present work, we focus on the second-order non-dissipative contributions to the current, $j^{\mu}_{(2)}$, and the energy-momentum tensor, $T^{\mu\nu}_{(2)}$. To isolate the effects relevant for our analysis, we retain only the possible second-order terms constructed from $a^{\mu}$ and $E^{\mu}$. We therefore begin by identifying all independent second-order structures built from these two vectors.
After some straightforward algebra, one finds that $\partial_{\mu}u_{\nu}=u_{\mu}a_{\nu}$,
$\partial_{\mu}a_{\nu}=a_{\mu}a_{\nu}-a^{2}u_{\mu}u_{\nu}$, and $\partial_{\mu}E_{\nu}=-(a\cdot E)u_{\mu}u_{\nu}+u^{\rho}\partial_{\mu}F_{\nu\rho}$.
Using these identities, we decompose $T_{\mu\nu}^{(2)}$ and $j_{\mu}^{(2)}$ as
\begin{eqnarray}
T^{\mu\nu}_{(2)} & = & e_{(2)}u^{\mu}u^{\nu}+P_{(2)}(g^{\mu\nu}-u^\mu u^\nu)\nonumber \\
 &  & +\kappa_1a^{\mu}a^{\nu}+\kappa_2E^{\mu}E^{\nu}+\kappa_3a^{(\mu}E^{\nu)}\nonumber \\
 &  & +\chi_{5}(u_{\lambda}\partial^{\mu}F^{\nu\lambda}+u_{\lambda}\partial^{\nu}F^{\mu\lambda}),\nonumber \\
j^{\mu}_{(2)} & = & n_{(2)}u^{\mu}+\xi_F \partial_{\lambda}F^{\lambda\mu},\label{eq:decomposition_01}
\end{eqnarray}
where the second-order corrections to the energy density $e_{(2)}$, pressure $P_{(2)}$, and number density $n_{(2)}$ are given by
\begin{eqnarray}
e_{(2)} & = & \lambda_1 a^{2}+\lambda_2 a\cdot E+\lambda_3 E^{2} \nonumber \\
&&+(\chi_{2}+\chi_{3})u^{\rho}\partial^{\lambda}F_{\lambda\rho},\nonumber \\
P_{(2)} & = & \overline{\lambda}_1 a^{2}+\overline{\lambda}_2 a\cdot E+\overline{\lambda}_3 E^{2}+\chi_{3}u^{\rho}\partial^{\lambda}F_{\lambda\rho},\nonumber \\
n_{(2)} & = & \xi_1a^{2}+\xi a\cdot E+\xi_2E^{2}.\label{eq:n_2}
\end{eqnarray}
The coefficients $\lambda_{i}$, $\overline{\lambda}_{i}$, $\kappa_{i}$, $\chi_{i}$ and $\xi_{i}$ are transport coefficients.

With the above decomposition, the conservation equation (\ref{conservation}) and the Weyl anomaly relation (\ref{eq:trace_hydro}) must be satisfied order by order. Interestingly, the second-order current conservation equation, $\partial^{\mu}j_{\mu}^{(2)}=0$, is automatically satisfied. By contrast, energy-momentum conservation imposes a set of constraints on the transport coefficients. Solving these constraints, we find that most of the transport coefficients introduced above are irrelevant to the Weyl anomaly.
We therefore focus on the relevant ones (see End Matter). We first find that
$\xi=T\partial_{T}\kappa_2$,
while $\chi_{3},\chi_{2},\kappa_3,\overline{\lambda}_2 $
can be expressed in terms of $\chi_{5}$.
In addition, the Weyl anomaly further imposes
\begin{equation}
2C=\lambda_3 +3\overline{\lambda}_3 +\kappa_2=-T\partial_{T}\overline{\lambda}_3 =\frac{1}{2}\xi.
\end{equation}
Substituting $\xi$ back into Eqs.~(\ref{eq:decomposition_01}) and (\ref{eq:n_2}), we obtain
\begin{equation}
j^{\mu}=4C(a\cdot E)u^{\mu},\label{eq:j_1}
\end{equation}
which means that the Weyl anomaly affects the conserved number density $n_{(2)}=u_\mu j^\mu$ at second order. 

The above analysis can also be extended to the case with nonvanishing magnetic fields. In a relativistic system, magnetic and electric fields generally appear together. To extract the magnetic-field contribution, it is convenient to rewrite the result in terms of the full field-strength tensor $F^{\mu\nu}$, which gives
\begin{equation}
j^{\mu}=-4CF^{\mu\nu}a_{\nu}.
\label{eq:j_2}
\end{equation}
We have also verified this result by keeping finite $B^{\mu}$ from the outset and repeating the same analysis within the hydrodynamic framework.


\vspace{0.5cm}
\emph{Rindler horizon and boundary QFT}---
We now show that the Weyl-anomaly-induced current in Eq.~(\ref{eq:j_2}) is independently reproduced within the framework of boundary quantum field theory (BQFT). In this approach, the effective field theory is organized as an expansion in inverse powers of the distance to the boundary, thereby incorporating quantum effects localized near it \cite{Deutsch:1978sc}.

\vspace{0.2cm}
\noindent
\textbf{Vacuum magnetization current from BQFT:}
Recently, a major development in BQFT was the discovery that the expectation value of the current in an external electromagnetic field $F^{\mu\nu}$ is related to the Weyl anomaly through the universal relation \cite{Chu:2018ksb,Chu:2018ntx, Chu:2019rod},
\begin{eqnarray}
j^{\mu}=\langle \hat{j}^\mu \rangle=4C\frac{F^{\mu\nu}n_{\nu}}{x}\,,\label{bqft}
\end{eqnarray}
where $n_{\nu}$ is the inward unit normal vector to the boundary and $x$ is the distance to the boundary. Here, “universal” means that the coefficient of the $1/x$ divergence is fixed by the Weyl anomaly coefficient, rather than by microscopic details or the choice of boundary condition.

\begin{figure}[t]
\includegraphics[width=1.0\linewidth]{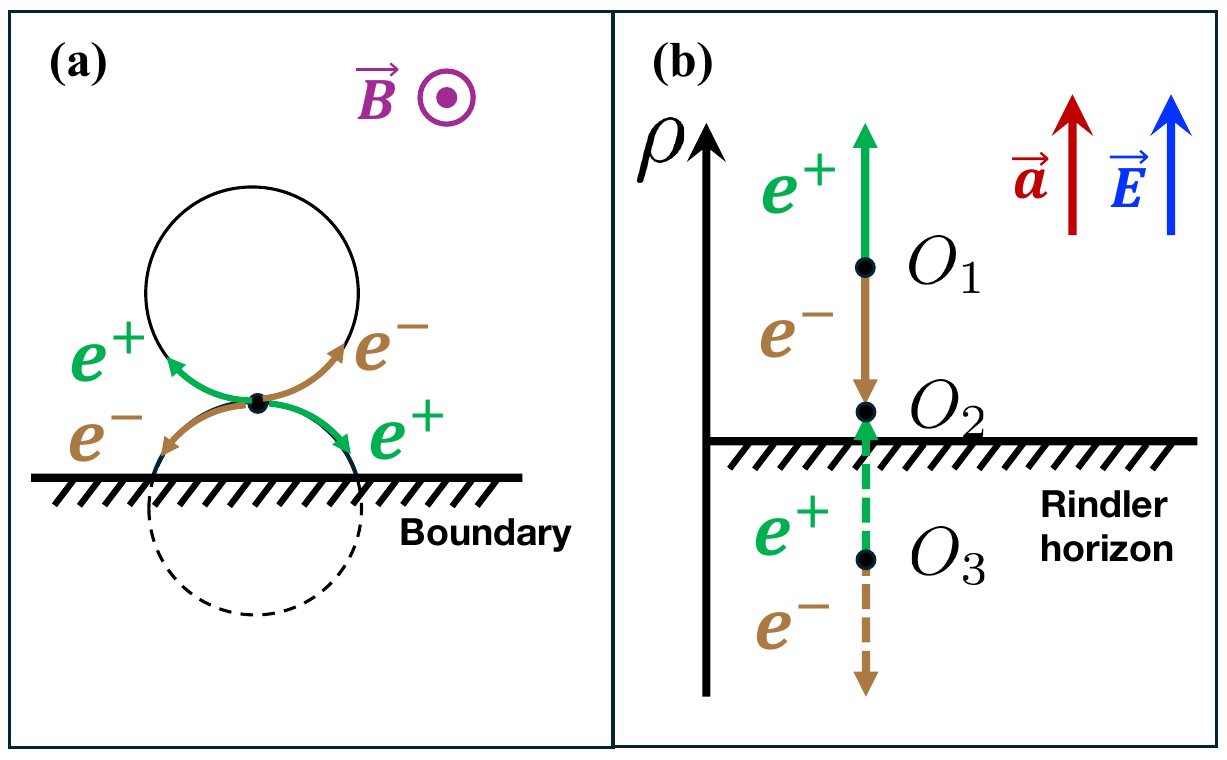}
\caption{Illustration of the Weyl-anomaly-induced current in Eq.~\eqref{eq:j_2}. Panel (a): vacuum magnetization current, following Ref.~\cite{Chu:2018ksb}. Panel (b): screening effect of the Rindler horizon induced by the electric field.}
\label{fig:1}
\end{figure}

To clarify the physical meaning of Eq.~(\ref{bqft}), let us consider the magnetic component of $F^{\mu\nu}$. Near the boundary, quantum fluctuations continually produce virtual charged pairs. In the presence of an external magnetic field, the particle and antiparticle are driven in opposite directions, tending to separate from each other near the boundary. As a consequence, the cancellation between their contributions becomes incomplete, and Eq.~(\ref{bqft}) yields a net current localized near the boundary; see also Fig.~\ref{fig:1}(a). Namely, there is no compensating current due to a particle-antiparticle pair born outside the system boundary. Therefore, this current can be interpreted as a vacuum magnetization current near the boundary, analogous to the Casimir effect, rather than as an ordinary transport current in matter.

\vspace{0.3cm}
\noindent
\textbf{Weyl-anomaly-induced current in an accelerated frame:} As mentioned above, the boundary plays a crucial role in Eq.~(\ref{bqft}). Without a boundary, the contributions from virtual particle-antiparticle pairs cancel, and no net effect can be observed. This naturally leads to the question: what happens if the horizon in an accelerated frame, namely the Rindler horizon, is treated as such a boundary?

To apply this framework to an accelerated observer, we consider Rindler spacetime \cite{Crispino:2007eb} corresponding to motion along the $z$-axis. The metric  is given by $ds^{2}=\rho^{2}d\theta^{2}-dx^{2}-dy^{2}-d\rho^{2}$,
where the Rindler coordinates $\rho$ and $\theta$ are related to the Minkowski coordinates by $z=\rho\cosh\theta$ and $t=\rho\sinh\theta$. The hypersurface $\rho=0$ corresponds to the horizon, since $g_{00}(\rho=0)=0$.
The radiation associated with this horizon is the well-known Unruh effect \cite{Unruh:1976db,Crispino:2007eb}.

Thus, $\rho$ corresponds to the distance from the horizon, $x$, which now plays the role of the boundary. Moreover, upon transforming the acceleration vector to Rindler coordinates, we obtain
\begin{eqnarray}
a_{\mu}=u^{\nu}\nabla_{\nu}u_{\mu}=-u^{\nu}\Gamma_{\nu\mu}^{\beta}u_{\beta}=(0,0,0,-1/\rho)\,.\label{hor}
\end{eqnarray}
Therefore, the vector $a_{\mu}$ points outward, namely away from the right Rindler wedge. The key point is that the horizon at $\rho=0$ is a degenerate surface whose normal vector is also tangent to the surface; equivalently, the normal to the Rindler horizon is null, $n_{0}^{\mu}n_{0,\mu}=0$. However, if we consider a ``stretched horizon'' \cite{Parikh:1997ma}, shifted by an infinitesimal distance to $\rho=\varepsilon$, then the normal to this shifted hypersurface, $n_{\mu}^{\varepsilon}$, is spacelike and satisfies $n_{\varepsilon}^{\mu}n_{\varepsilon,\mu}=-1$. The inward unit normal vector to the stretched horizon is then given by $n_{\mu}^{\varepsilon}=(0,0,0,1)$. Therefore, the acceleration vector can be expressed in terms of the normal to the stretched horizon and the distance to it as
\begin{eqnarray}
\frac{n_{\mu}^{\varepsilon}}{x}=-a_{\mu}\,.
\label{sub}
\end{eqnarray}
Assuming that the ``stretched horizon'' can be treated analogously to the boundary in Eq.~(\ref{bqft}), we expect a similar effect in Rindler space, with the substitution in Eq.~(\ref{sub}). We then obtain our main result,
\begin{equation}
j^{\mu}=4C[(a\cdot E)u^{\mu}-\epsilon^{\mu\nu\alpha\beta}u_{\nu}B_{\alpha}a_{\beta}],
\label{eq:main_results}
\end{equation}
which is exactly the same as the result obtained from the hydrodynamic analysis in Eq.~(\ref{eq:j_2}). Although Eq.~(\ref{sub}) and the main result (\ref{eq:main_results}) are derived in Rindler space, they are written entirely in terms of covariant quantities and therefore apply in any reference frame.

\vspace{0.2cm}
\noindent
\textbf{Screening effects induced by electric fields:} 
Similar to the vacuum magnetization current, the same interpretation applies to Eq.~(\ref{sub}) and to the first term, proportional to $E^{\mu}$, in Eq.~(\ref{eq:main_results}), since they are independent of temperature and density. Related vacuum effects in accelerated media have long been discussed in the literature \cite{Dowker:1987pk,Candelas:1978gg,Prokhorov:2019yft}. We illustrate a simple qualitative picture for the current in Eq.~\eqref{eq:j_1} in Fig.~\ref{fig:1}(b), following the spirit of Refs.~\cite{Chu:2018ksb,Chernodub:2021nff}.

Returning to the accelerated frame and Rindler space, consider an
electric field aligned with the acceleration. The quantum vacuum is
filled with fluctuations which, for an accelerated observer, appear
as thermal radiation, that is the essence of the Unruh effect. Let
us consider the vacuum fluctuation of an electron and a positron at
a point $O_{1}$ near the Rindler horizon in an external electric
field. Under the influence of an electric field, particles with different
charges will begin to move in opposite directions. As a result, a
virtual electron with a negative charge from point $O_{1}$ will come
to point $O_{2}$. If there were no horizon, then at point $O_{2}$
it would meet a positron born at point $O_{3}$, and the charge at
point $O_{2}$ would be zero. However, due to the horizon, point $O_{3}$
is inaccessible. As a result, a negative electric charge is created
in $O_{2}$. This corresponds to a first term in Eq. (\ref{eq:main_results}). Note
that the corresponding charge screens the electric field, creating
a field in the opposite direction. Thus, due to the current (\ref{eq:main_results}),
the Rindler horizon is screened.


\vspace{0.5cm}
\emph{Possible physical implications}---Having established the Weyl-anomaly-induced current through two independent derivations, we now place this result in the broader context of anomalous transport. Without loss of generality, we work in the local rest frame of the fluid, $u^{\mu}=(1,\boldsymbol{0})$, in which our main result, Eq.~(\ref{eq:main_results}), reduces to
\begin{equation}
j^{0}=-4C(\mathbf{E}\cdot\mathbf{a}),\;\mathbf{j}=4C\mathbf{a}\times\mathbf{B}.\label{eq:j_3}
\end{equation}

\begin{figure}[t]
\includegraphics[width=0.8\linewidth]{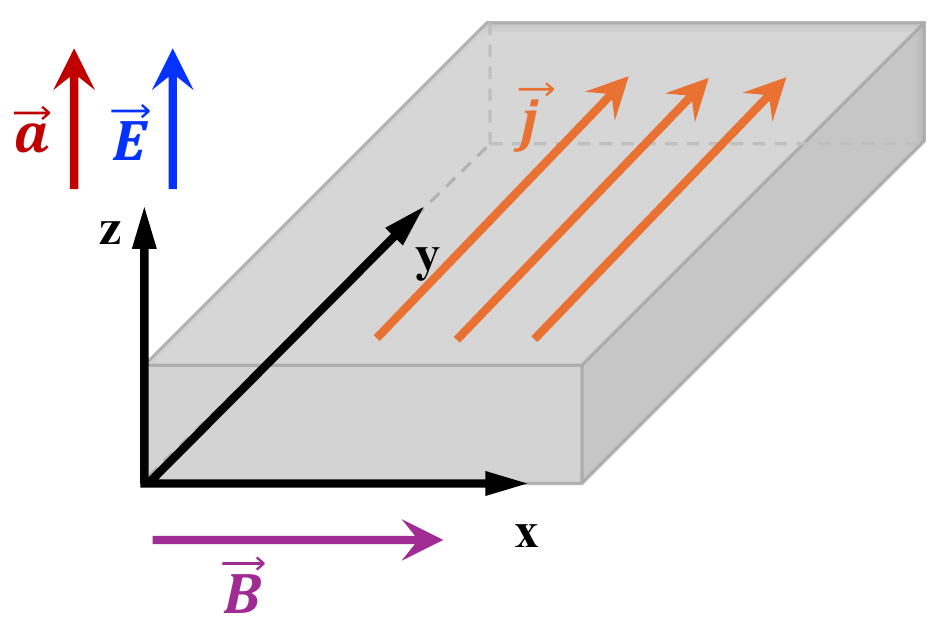}
\caption{Illustration of the Weyl-anomaly-induced current in Eq.~\eqref{eq:j_4} at global equilibrium in the local rest frame.}
\label{fig:2}
\end{figure}

The simplest case is to consider the system at global equilibrium, as in the hydrodynamic framework. In this case, the Killing condition in Eq.~(\ref{eq:Killing_eq_02}) implies $\partial_{t}\overline{\mu}=0$ and $\mathbf{E}=T\boldsymbol{\nabla}\overline{\mu}$. This means that, at global equilibrium, the force induced by the external electric field is balanced by the diffusive contribution associated with $\boldsymbol{\nabla}\overline{\mu}$. From Eq.~(\ref{eq:a_def}), we further obtain $a^{\mu}=(0,T^{-1}\boldsymbol{\nabla}T)$ together with $\partial_{t}T=0$, namely that the spatial inhomogeneity of the temperature induces the local acceleration. Equation~(\ref{eq:j_3}) then reduces to
\begin{eqnarray}
j^{0} & = & -4C\boldsymbol{\nabla}\overline{\mu}\cdot\boldsymbol{\nabla}T,\;\mathbf{j}=4CT^{-1}\boldsymbol{\nabla}T\times\mathbf{B}.
\label{eq:j_4}
\end{eqnarray}
The combination of the local acceleration induced by $\boldsymbol{\nabla}T$ and the diffusive contribution associated with $\boldsymbol{\nabla}\overline{\mu}$ leads to an additional accumulation of the local charge density. Meanwhile, the $\boldsymbol{\nabla}T\times\mathbf{B}$ contribution to the current has the same tensor structure as a Nernst-like \cite{Kavokin:2020NernstCorbino, Ferreiros:2017AnomalousNernstTiltedWeyl}, or more generally thermomagnetic Hall-like, transverse current \cite{Ferreiros:2017AnomalousNernstTiltedWeyl, Uchida:2022TransverseThermoelectrics, Sekine:2020QuantumKineticThermoelectric}, but its microscopic origin is different. In our case, the coefficient is fixed by the Weyl anomaly rather than by ordinary quasiparticle transport; see also Fig.~\ref{fig:2} for an illustration. This may also be relevant to condensed matter systems, especially Weyl semimetals, where conformal-anomaly-related effects have already been discussed and observed \cite{Chernodub:2021nff}.

Beyond global equilibrium, the acceleration $\mathbf{a}$ can arise from more general sources, so this effect may operate in a wider class of systems. A natural example is relativistic heavy-ion collisions, where strong electromagnetic fields \cite{Bzdak:2011yy, Deng:2012pc, Pu:2016ayh}, strong vorticity \cite{STAR:2017ckg,Deng:2016gyh,Yi:2026rbz}, and other sources of fluid acceleration \cite{Prokhorov:2025vak,Karpenko:2018erl} coexist. The resulting Weyl-anomaly-induced vector current may affect the evolution of the quark-gluon plasma and contribute to anomaly-driven charge transport and charge separation.
The universality of the Weyl anomaly suggests that this transport mechanism may also have implications beyond heavy-ion collisions. In cosmology, the trace anomaly can induce dynamical effects in the early universe \cite{Chernodub:2016lbo}. During hot, rapidly evolving stages with strong electromagnetic fields and large local accelerations, such as reheating or near phase transitions, the equilibrium current may contribute to anomaly-driven charge transport relevant to charge- and baryon-asymmetry generation.

\vspace{0.5cm}
\emph{Summary}---We have identified a Weyl-anomaly-induced current, Eq.~\eqref{eq:main_results}, from second-order relativistic hydrodynamics at global equilibrium, and independently reproduced the same result from BQFT as a nontrivial boundary-induced effect. 
From the BQFT perspective, the magnetic- and electric-field-induced Weyl-anomaly currents correspond to vacuum magnetization and screening effects near the boundary.
In the local rest frame, the electric-field contribution leads to an additional accumulation of charge density, while the magnetic-field contribution generates an additional current. At global equilibrium, the magnetic-field-induced current has the same tensor structure as a Nernst-like, or more generally thermomagnetic Hall-like, response.

These results point to a new class of anomaly-induced transport governed by the trace anomaly, with the corresponding transport coefficient fixed by the Weyl anomaly. Our findings suggest broader implications for anomaly-induced transport in many-body systems.

\vspace{0.5cm}
{\bf Acknowledgments}
This work is supported in part by the National Key Research and Development Program of China under Contract No. 2022YFA1605500, by the Chinese Academy of Sciences (CAS) under Grant No. YSBR-088 and by National Natural Science Foundation of China (NSFC) under Grants Nos.~12575147, 12321005, 12175123, and 12135011. The work of G. Yu. Prokhorov and V. I. Zakharov was supported in part by Russian Science Foundation Grant No. 25-22-00887.

\bibliography{lit}
\bibliographystyle{apsrev4-1}

\onecolumngrid
\section*{End Matter}
\twocolumngrid
\emph{Mapping second-order hydrodynamics to the Weyl anomaly}---
From energy-momentum conservation, we obtain the set of constraints that determine the transport coefficients,
\begin{eqnarray}
\xi & = & T\partial_{T}\kappa_2-T^{-1}\partial_{\overline{\mu}}(\overline{\lambda}_2 +\kappa_3+\chi_{5})\,,\label{con-1}\\
0 & = & \lambda_3 -\overline{\lambda}_3 -\kappa_2+T\partial_{T}\overline{\lambda}_3 \nonumber \\
&&-T^{-1}\partial_{\overline{\mu}}(\kappa_3-\chi_{5})\,,\label{con-4}\\
0 & = & \overline{\lambda}_2 +\kappa_3-\chi_{5}+2T\partial_{T}\chi_{5}\,,\label{con-5}\\
0 & = & \kappa_3+\chi_{2}-\chi_{5}+T\partial_{T}\chi_{3}\,,\label{con-6}\\
0 & = & 2\overline{\lambda}_3 +\kappa_2-2T^{-1}\partial_{\overline{\mu}}\chi_{5}\,,\label{con-3}\\
0 & = & \chi_{3}+2\chi_{5}\,.\label{con-8}
\end{eqnarray}
In addition, Weyl anomaly imposes the further conditions 
\begin{eqnarray}
2C & = & \lambda_3 +3\overline{\lambda}_3 +\kappa_2\,,\label{con-2-1}\\
0 & = & \chi_{2}+4\chi_{3}+2\chi_{5}\,.\label{con-7}
\end{eqnarray}
From Eqs.(\ref{con-8}) and (\ref{con-7}), we deduce 
\begin{eqnarray}
\chi_{3}=-2\chi_{5}\,,\ \chi_{2}=6\chi_{5}\,.\label{relation-1}
\end{eqnarray}
The introduced coefficients $\lambda, \kappa ...$ have dimensions by mass -1, 0, 1 or 2. Due to this, one should expect, for example, that $\chi_{5}=b(\overline{\mu})T$, and similarly for other coefficients, which can be considered as a choice of solution motivated by dimensional counting.
Using the results obtained above and substituting these relations into Eqs.~(\ref{con-5}) and (\ref{con-6}), we find
\begin{eqnarray}
\kappa_3=-3\chi_{5},\ \ \ \ \overline{\lambda}_2 =2\chi_{5}\,.\label{relation-2}
\end{eqnarray}
With these results, Eqs.~(\ref{con-1}) and (\ref{con-4}) simplify
to, 
\begin{eqnarray}
\xi & = & T\partial_{T}\kappa_2\,,\label{con-1-a}\\
0 & = & \lambda_3 -\overline{\lambda}_3 -\kappa_2+T\partial_{T}\overline{\lambda}_3 \nonumber \\
&&+4T^{-1}\partial_{\overline{\mu}}\chi_{5}\,.\label{con-4-a}
\end{eqnarray}
Combining Eqs.~(\ref{con-3}) and (\ref{con-4-a}), we obtain 
\begin{eqnarray}
0 & = & \lambda_3 +3\overline{\lambda}_3 +\kappa_2+T\partial_{T}\overline{\lambda}_3 .
\end{eqnarray}
Using the anomaly relation (\ref{con-2-1}), we get 
\begin{eqnarray}
2C & = & -T\partial_{T}\overline{\lambda}_3 .\label{relation-0}
\end{eqnarray}
This result shows that for non-zero constant $C$, the coefficient $\overline{\lambda}_3$ cannot be naively \
assumed to have a $T^0$   dependence on temperature from dimensional counting; 
instead, it exhibits a logarithmic dependence.
Taking the partial derivative of Eq.~(\ref{con-3}) with respect to $T$, we further obtain 
\begin{eqnarray}
\partial_{T}\kappa_2 & = & -2\partial_{T}\overline{\lambda}_3 \,.\label{relations-3}
\end{eqnarray}
That gives us Eq.~\eqref{eq:j_1}.

\vspace{-1.0em}

\end{document}